\documentstyle[onecolumn,psfig,float]{mn}

\newcommand{\be}{\begin{equation}}
\newcommand{\ee}{\end{equation}}
\newcommand{\bea}{\begin{eqnarray}}
\newcommand{\eea}{\end{eqnarray}}

\newcommand{\dl}{\delta}

\newcommand{\eps}{\epsilon}

\newcommand{\th}{\theta}

\newcommand{\Lm}{\Lambda}

\newcommand{\ph}{\phi}
\newcommand{\Ph}{\Phi}

\newcommand{\om}{\omega}

\newcommand{\nn}{\nonumber}

\title[Torsional Oscillations of Magnetized Relativistic Stars]
{Torsional Oscillations of Magnetized Relativistic Stars}

\author[Neophytos Messios, 
        Demetrios B. Papadopoulos and 
        Nikolaos Stergioulas] 
       {Neophytos Messios, Demetrios B. Papadopoulos and
        Nikolaos Stergioulas \\
        Department of Physics, 
        Section of Astrophysics, Astronomy and Mechanics,
        Aristotle University of Thessaloniki, \\
        54006 Thessaloniki, Greece}

\date{}
\pagerange{}
\pubyear{} 

\begin{document}

\maketitle

\label{firstpage}    

\begin{abstract}
  
  Strong magnetic fields in relativistic stars can be a cause of crust
  fracturing, resulting in the excitation of global torsional
  oscillations. Such oscillations could become observable in
  gravitational waves or in high-energy radiation, thus becoming a
  tool for probing the equation of state of relativistic stars. As the
  eigenfrequency of torsional oscillation modes is affected by the
  presence of a strong magnetic field, we study torsional modes in
  magnetized relativistic stars. We derive the linearized perturbation
  equations that govern torsional oscillations coupled to the
  oscillations of a magnetic field, when variations in the metric are
  neglected (Cowling approximation). The oscillations are described by
  a single two-dimensional wave equation, which can be solved as a
  boundary value problem to obtain eigenfrequencies.  We find that in
  the non-magnetized case, typical oscillation periods of the
  fundamental l=2 torsional modes can be nearly a factor of two larger
  for relativistic stars than previously computed in the Newtonian
  limit. For magnetized stars, we show that the influence of the
  magnetic field is highly dependent on the assumed magnetic field
  configuration and simple estimates obtained previously in the
  literature cannot be used for identifying normal modes
  observationally.

\end{abstract}


\section{Introduction}

Although the interior composition of relativistic stars is currently
still very uncertain, the properties of their solid crust have been
studied extensively (Ruderman 1968, see Pethick \& Ravenhall 1999 for a
recent review).  It has been suggested that the presence of a strong
magnetic field in a secularly evolving star can cause the crust to
fracture, exciting shear waves and leading to the phenomenon of soft
gamma repeaters (SGRs) (Cheng et al. 1996, see Thompson 2000 for a
recent review of SGRs). In such a scenario, the oscillation modes that
appear most favored for excitation are the low-order torsional
oscillations of the crust (Duncan 1998). Torsional oscillations differ
from spheroidal shear oscillations, in that they are predominantly
divergence-free, toroidal velocity oscillations, accompanied by only
small density oscillations in the presence of rotation or a magnetic
field. Although torsional oscillations of neutron star crusts have a
strong potential for observational detection, the study of these modes
in the literature has been limited, so far, to a few representative
models of the neutron star equilibrium and crust structure, while the
magnetic field has only been taken into account in an approximate,
Newtonian framework.

Torsional modes may be favored for excitation (compared to spheroidal
shear modes) during a starquake, because the restoring force is mainly
due to the relatively weak Coulomb forces of the crustal ions (thus
requiring much less energy than compressional oscillations) and
because the lowest-order torsional oscillation frequency has a 
relatively long period, which implies a slow damping rate (Duncan
1998). Although no oscillation modes in neutron stars have been
detected to date, there is evidence for a 23 ms periodicity in the
initial pulse of the 1979 March 5 event (Barat et al. 1983). If
confirmed, then torsional modes could become the first normal modes in
neutron stars to be detected. As the frequencies of the torsional
modes are significantly affected in the presence of a strong magnetic 
field, accurate frequencies for these modes for various magnetic
field strengths and magnetic field configurations are needed, if SGRs 
are indeed strongly magnetized compact stars.

In the Newtonian framework, torsional modes were first studied by
Alterman, Jarosh \& Pekeris (1959), applied to oscillations of the
Earth's solid crust. That torsional modes could also be excited in
neutron star crusts was first proposed by Ruderman (1968) and studied
by Hansen \& Cioffi (1980), Van Horn (1980), McDermott et al. 1985
and McDermott, Van Horn \& Hansen (1988). Torsional modes of slowly
rotating neutron stars have been studied by Lee \& Strohmayer (1996).
In general relativity, the theory of torsional oscillations was
developed by Schumaker \& Thorne (1983), applying the
general-relativistic theory of elasticity by Carter \&
Quintana (1972) (for a discussion, see Priou 1992 and references
therein). The only numerical computation of torsional modes for
realistic neutron star crusts in general relativity has been presented
by Leins (1994).

In the present work, we initiate the study of torsional oscillations
in relativistic stars possessing a strong magnetic field. Our
theoretical description is based on the study of wave propagation in
hydromagnetic media in general relativity by Papadopoulos \& Esposito
(1982). Previously, the magnetic field effect on the torsional modes
has been studied in the Newtonian limit by Carroll et al. (1986) and
by Nasiri \& Sobuti (1989) (see also Duncan, 1998).  Here, we derive
the linear perturbation equations governing torsional oscillations in
a magnetized star in the relativistic Cowling approximation
(Cowling 1941; McDermott, Van Horn \& Scholl, 1983), neglecting the deformation of
the equilibrium structure, due to the presence of a magnetic field.
For a general axisymmetric magnetic field configuration, the
perturbation equations are two-dimensional.  Simplified,
one-dimensional equations are derived for a special case of the
magnetic field configuration. In future work, we plan to include the
deformation of the neutron star structure and study more general
magnetic field configurations by solving the two-dimensional
eigenvalue problem for various magnetic field configurations.

\section{The Equilibrium Configuration}

The equilibrium structure of a magnetized relativistic star is
non-spherical, due to the deformation induced by magnetic field
stresses.  Self-consistent models of relativistic stars with a strong
magnetic field have been constructed by Bocquet et al. (1995) (in the
case of rapidly rotating stars) and Gupta et al. (1998) (in the
slow-rotation limit). The spherical structure of a nonrotating
relativistic star is distorted significantly, only when the magnetic
field becomes very large, exceeding $\sim 10^{14}$G (for a magnetic
dipole). Although this distortion will certainly modify the
eigenfunctions of various oscillation modes, we expect this effect to
be smaller than the direct influence of the magnetic field on the
eigenfunctions and eigenfrequencies of oscillations, unless the
magnetic field becomes exceedingly large. Thus, here we will neglect
the distortion induced by the magnetic field and will assume the
equilibrium structure to be that of a spherical relativistic star, the
space-time of which is described by the metric \be ds^{2}=-e^{2\Ph(r)
  }dt^{2} +e^{2\Lm(r) }dr^{2}+r^{2} \left(d{\th} ^{2}+ \sin^{2}{\th} d\ph ^{2}
\right), \ee where ${\Ph }(r)$ and ${\Lm}(r)$ are functions of the
radial coordinate only.  The equilibrium structure is obtained in
full general relativity (see e.g.  Schumaker \& Thorne 1983),
assuming a perfect-fluid equation of state.  Although our purpose is
to study oscillations in the solid crust of the star, the bulk
properties and local pressure and density profiles in the crust can be
obtained accurately, by assuming a perfect fluid.  We assume, further,
that the crust has an isotropic shear modulus.  The shear modulus only
appears in the perturbed configuration.

In order to derive the perturbation equations for a magnetized relativistic
star, the equilibrium stress-energy tensor is assumed to consist of two
parts:
\be
\label{stress}
T^{ab}=T^{ab({\rm pf})}+T^{ab({\rm M})}, \label{Tab}\ee where \be
T^{ab({\rm pf})} =\left(\eps+p \right)u^{a}u^{b} +pg^{ab}, \ee and \be
T^{ab({\rm M})} = H^{2}u^{a}u^{b}+\frac{H^{2}}{2}g^{ab}-H^{a}H^{b} \ee
(see Papadopoulos \& Esposito, 1982). Above, $\epsilon$ and $p$ are the energy
density and pressure of the fluid and $u^a$ is the four-velocity of
fluid elements \be u^{a}=\left[e^{-\Phi(r)},0,0,0\right], \ee (with $u_a
u^a=-1$), while $H^{a}$ are the components of the magnetic field (with
$H^2=g_{ab}H^aH^b$). The solid crust is assumed to be isotropic and
perfectly elastic, so that in equilibrium there is no shear stress
contribution in (\ref{stress}).  Here, and throughout the paper, we
are using the same units as in Papadopoulos \& Esposito (1982), i.e.
we set $c=G=1$ and normalize the magnetic field by 
multiplying $H^{a}$ by $\sqrt{4\pi}$, unless otherwise noted.

The equations of motion of the fluid are obtained by the projection of
the conservation of the stress-energy tensor onto the hypersurface normal
to $u^a$
\begin{equation}
h^c{}_a T^{ab}{}_{;b}=0,
\end{equation}
(where $h^c{}_a=g^c{}_a+u^cu_a$ and ``;'' denotes the covariant derivative
compatible with the equilibrium metric $g_{ab}$), which reads
\begin{equation}
\left(\eps+p+H^{2}\right) u^{b}u^{a}{}_{;b} = 
-\left(p+\frac{H^{2}}{2}\right)_{;b}
 h^{ab}+h^{a}{}_{c} \left(H^{c}H^{b} \right)_{;b}. \label{Euler}
\end{equation}
For perfectly conducting hydromagnetic media, Maxwell's equations 
$F^{ab}{}_{;b}=0$ (where $F^{ab}$ is the electromagnetic tensor) take the form
\begin{equation}
\left( u^aH^b-u^bH^a \right)_{;a}=0,
\end{equation}
which leads to the following form of the magnetic induction equation
\be
{H}^a{}_{;b}u^b=\left(\sigma^{a}{}_{b}+\omega^{a}{}_{b}
-\frac{2}{3}\delta^{a}{}_{b}\Theta  \right)H^b+ H^b u_{b;c}u^c u^a,
\label{Maxwell} \ee
where $\Theta =u^{b}{}_{;b}$ is the expansion of the fluid, $\omega^a{}_b$ is
the twist tensor and $\sigma^a{}_b$ is the shear tensor. We assume a zero
net electrical surface charge, so that there is no equilibrium 
electric field. The assumption of infinite conductivity is justified
as the magnetic field diffusion timescale is several orders of magnitude
larger than the typical timescale for torsional oscillations.

\section{The Perturbed Configuration}

The various oscillation modes that can be present in a star are
studied by linearizing the equations governing the equilibrium
configuration and assuming a harmonic time-dependence. Here, we are
only interested in the torsional modes of the crust, which do not
generate significant variations in the gravitational field of the
star. In addition, the low-order quadrupole torsional mode has a
long period of the order of 20-40ms. Thus, such an oscillation can be
described well by the slow-motion approximation (Schumaker \& Thorne
1983). It follows that one can describe the low-order torsional
oscillations of the crust by neglecting the variations in the metric,
i.e. by setting $\delta g_{ab}=0$, which is usually called the
relativistic Cowling approximation (McDermott et al. 1983) (this
approximation was also used in the description of hydromagnetic waves
in general relativity by Papadopoulos \& Esposito, 1982). This
approximation yields the real part of the eigenfrequencies with a
typical accuracy of a few percent, as can be deduced from the results
in Leins (1994) and we will employ it throughout this paper. Once the
eigenfunction and the real part of the eigenfrequency are computed,
estimates for the damping rate of torsional oscillations due to
gravitational and electromagnetic wave emission and due to viscous
dissipation can be obtained in a perturbative way (see e.g. McDermott
et al.  1988), since the imaginary part of the eigenfrequency is much
smaller than the real part.

The linearized version of the equations of motion (\ref{Euler})  is 
\begin{eqnarray}
\label{ls}
\left(\epsilon +p+H^2\right) \dl u^{a}{}_{;b}u^{b} &=& 
-\left(\dl \eps +\dl p 
 + 2 H^c \delta H_c \right) u^{a}{}_{;b}u^{b}  
+\left(u^{a}\dl u_c  +\dl u^{a}u_{c} \right) 
\left[H^{c}H^{b}-g^{cb} \left( p+\frac{H^2}{2} \right) \right]_{;b}  \nn \\
&&  - \left(\epsilon +p+H^2\right) u^{a}{}_{;b}\dl u^{b} 
 +h^a{}_c \left[H^c\delta H^b+\delta H^c H^b 
- g^{cb} \left(\delta p + H^d \delta H_d \right) \right]_{;b}  \nn \\
&& -h^{a}{}_{c}\dl T^{cb}{}_{;b}^{({\rm S})}. \label{pEuler}
\end{eqnarray} 
In (\ref{pEuler}), the perturbation in the shear stress tensor is \be
\dl T^{({\rm S})}_{ab}=-2\mu \delta S_{ab}, \ee where $\delta S_{ab}$ is the
perturbation in the strain tensor and $\mu$ is the isotropic shear
modulus.  The linearized version of the magnetic induction equation
(\ref{Maxwell}) is \bea (\delta H^a)_{;c} u^c &=& -H^{a}{}_{;c} \delta u^c+
h^{ac} H^d(\delta u_c )_{;d} +h^{ac} u_{c;d} \delta H^d +(u^a\delta u^c+u^c \delta u^a)H^d
u_{c;d} -\theta \delta H^a -H^a \delta \theta \nn \\ && -u^a H^c \left[ (\delta u_c)_{;b}
  u^b+u_{c;b} \delta u^b \right] - \left( H^c \delta u^a +u^a \delta H^c \right)
u_{c;b} u^b. \label{pM} \eea Equations (\ref{pEuler}) and (\ref{pM})
are the complete set of equations that govern the perturbations of
magnetized relativistic stars in the Cowling approximation.

\section{The Eigenvalue Problem}

When the distortion of the equilibrium structure, due to the magnetic
field, is ignored, the spherical symmetry of the unperturbed star
allows for the oscillations to be decoupled into modes of definite
spherical-harmonic indices $(l,m)$ and definite parity. Here, we
investigate pure torsional oscillations, which are the normal modes of
odd (magnetic-type) parity, $\pi=(-1)^{l+1}$ (Regge \& Wheeler, 1957).
In spherical symmetry, modes with fixed $l$ but different $m$ 
yield the same frequency, thus, we will specialize to the case of
$m=0$ only. This means that, we will not study (at this point) the
mode-splitting caused by the magnetic field.

When expanded in vector spherical harmonics of definite $l$ and $m=0$,
the odd-parity perturbation in the four-velocity can be written as
(see Schumaker \& Thorne, 1983) 
\bea
\dl  u^{\th} &=& 0 \\
\dl u^{\ph} &=& e^{-\Ph}\frac{\partial Y(r,t)}{\partial t} b^{\ph}, 
\eea 
where
\be
b^{\Phi}=\frac{1}{\sin {\th}} \frac{d P_{l}(\cos{\th})}{d \theta},
\ee
and $Y(r,t)$ is the angular displacement of the oscillating stellar
material and $P_{l}(\cos {\th})$ is the Legendre polynomial of order
$l$. For odd-parity perturbations in a spherical background: $\delta
u^{t}=\dl u^{r}=\dl \epsilon =0$.
 
We assume a harmonic time-dependence of all perturbed variables, as in
$Y(r,t)=Y(r)e^{i\om t}$, where $\omega$ is the mode frequency (from now
on we drop the time dependence from all perturbed variables). The
$\ph-$component of the perturbed equations of motion (\ref{pEuler})
becomes
\bea
i \omega \left(\epsilon +p+H^2 \right) e^{-\Phi} \dl u^{\phi} &=& 
\delta H^{r} \left[H^{\phi}{}_{,r}+H^{\phi} \left(\Phi_{,r}+\Lambda_{,r}
+\frac{4}{r} \right)- \frac{e^{2 \Lambda} H^r_{,\phi}}{r^2 \sin^2\theta} \right] 
 +\delta H^{\theta} \left[H^{\phi}{}_{,\theta} +3 \cot\theta 
H^{\phi}- \frac{ H^\theta_{,\phi}}{\sin^2\theta} \right] \nn \\
&& +\delta H^{\phi} \left[ 2\cot\theta H^\theta +\frac{2}{r}
H^{r} + H^r ( \Lambda_{,r} + \Phi_{,r}) \right] +H^{r}\delta H^{\phi}{}_{,r} \nn \\ 
&& +H^{\phi} \left[\delta H^{r}{}_{,r}+\delta H^{\theta}{}_{,\theta}
+\delta H^{\phi}{}_{,\phi} \right] +H^{\theta}\delta H^{\phi}{}_{,\theta} 
 +H^{\phi}\delta H^{\phi}{}_{,\phi} - \frac{H^r\delta H_{r,\phi}}{r^2 \sin^2\theta} 
-\frac{H^{\theta}\delta H_{\theta,\phi}}{r^2 \sin^2\theta} \nn \\ 
&& -\dl T^{r \ph  }{}_{,r}^{({\rm S})} -\dl T^{\theta \ph}{}_{,{\th}}^{({\rm S})}
- \left(\frac{4}{r}+\Ph_{,r}+\Lm_{,r} \right) \delta T^{r \ph ({\rm S})}- 3\cot {\th} 
\delta T^{{\th}\ph ({\rm S})},
\label{du^phi}
\eea  
where
\be
\dl T^{({\rm S})}_{r \ph}=-\mu r^{2} \sin^{2}{\th} 
e^{-\Ph}\frac{\partial Y}{\partial r}b^{\Ph}, \label{dTrphi}
\ee
and 
\be
\dl T^{({\rm S})}_{ \theta \phi}=-\mu r^{2} 
\sin^{2}{\th} e^{-\Ph} Y\frac{\partial b^{\Ph}}{\partial {\th}}, \label{dTthetaphi}
\ee
are the only non-vanishing components of the perturbed shear stress
tensor. The oscillating magnetic field components are given algebraically
as
\bea
\delta H^r &=& 0, \label{dHr} \\
\delta H^\theta &=& 0,  \label{dHtheta} \\
\delta H^\phi &=& \frac{e^\Phi}{i\omega}
 \left( H^r \delta u^\phi{}_{,r} +
 H^\theta \delta u^\phi{}_{,\theta} \right). \label{dHphi}
\eea
Equation (\ref{du^phi}) represents a two-dimensional boundary value 
problem for the eigenfrequency $\omega$ and the eigenfunction $Y(r)$. 
If we substitute  (\ref{dTrphi})-(\ref{dHphi}) in
(\ref{du^phi}), we arrive at our final expression 
\bea -\left(\epsilon
  +p+H^2\right) \omega^2 Y &=& \mu e^{2(\Ph-\Lm)}Y_{,rr}-(l+2)(l-1)\mu \frac{
  e^{2\Ph}}{r^{2}} Y +\left[ \left(\frac{4}{r}+\Ph_{,r}-\Lm_{,r}
  \right)\mu
  +\mu_{,r} \right] Y_{,r}e^{2(\Ph -\Lm)} \nn \\
&&+\frac{e^{2\Phi}}{ b^{\Phi} } \Biggl\{ b^{\Phi} \Bigl[H^r \left(
  2H^{\theta}\cot\theta+\frac{2}{r}H^r + H^r(\Lambda_{,r}+\Phi_{,r})\right) 
\left(-\Phi_{,r}Y+Y_{,r} \right)
+H^r H_{,r}^r \left(-\Phi_{,r} Y+Y_{,r} \right) \nn \\
&& +H^{\theta}H_{,\theta}^r \left(-\Phi_{,r} Y+Y_{,r} \right)
+(H^r)^2 \left(-\Phi_{,r} Y+Y_{,r} \right)_{,r} \Bigr]\nn \\
&&+b_{,\theta}^{\Phi} \Bigl[ 2H^{\theta}Y \left(H^{\theta}\cot\theta+\frac{1}{r}H^r \right)
+H^r \left( H_{,r}^{\theta}Y+H^{\theta} Y_{,r} \right)
+H^{\theta}H^r \left(-\Phi_{,r} Y+Y_{,r} \right) \nn \\
&& +H^{\theta}H_{,\theta}^{\theta}Y \Bigr] +b_{,\theta\theta}^{\Phi}(H^{\theta})^2Y \Biggr\},
\label{final}
\eea 
 In the absence of a magnetic field, separation of variables allows the above
equation to be reduced to a one-dimensional equation with respect to
the radial coordinate. This is not possible for a general magnetic
field configuration. However, as we will show in Section \ref{special},
there exist special cases of the magnetic field, for which the problem
becomes one-dimensional. As the numerical solution of the
two-dimensional problem is not a trivial task, such one-dimensional
cases are very useful for obtaining initial order-of-magnitude estimates for
the influence of a strong magnetic field on the torsional oscillations.

 \section{Analytic estimates}
\label{simple}

If one assumes a uniform density star with uniform shear modulus $\mu$
in the Newtonian limit, one can derive a simple analytic estimate for
the influence of the magnetic field on the period of torsional
oscillations.  In the nonmagnetized case, the period of torsional
oscillations is given analytically as
\begin{equation}
P=\frac{2\pi R}{x_nv_s},
\end{equation}
where $v_s=\sqrt{\mu / \rho}$ is the speed of sound, $\rho$ is the
density of the crust, $R$ is the star's radius and $x_n$ is a constant
(see Schumaker and Thorne 1983). In the presence of a magnetic field $B$,
the density $\rho$ is replaced by $\rho+B^2/4 \pi$. If one also assumes that the 
shear modulus $\mu$ is augmented by the magnetic field tension $B^2/4\pi$,
one easily obtains
\begin{equation}
P = P_0\sqrt{\frac{1+v_A^2}{1+(B/B_\mu)^2}},
\label{P1}
\end{equation} 
where $v_A=B^2/4 \pi \rho$ is the Alfv{\'e}n speed, $B_\mu = (4 \pi \mu )^{1/2}$ 
and $P_0$ is the oscillation period for a nonmagnetized star. This
can be rewritten as
\begin{equation}
P = P_0\sqrt{\frac{1+v_s^2(B/B_\mu)^2}{1+(B/B_\mu)^2}} \simeq P_0[1+(B/B_\mu)^2]^{-1/2},
\label{P2}
\end{equation} 
when $v_s< < 1$ (which was also considered by Duncan, 1998).
The above estimate implies that the main influence of the magnetic
field on the oscillation period is through the magnetic field tension.
Based on a power-law fit to the deep-crust equilibrium composition
by Negele and Vautherin (1973), Duncan (1998) 
estimates $B_\mu=4 \times 10^{15} \rho^{0.4}_{14}$G, where $\rho_{14}=\rho/10^{14}{\rm gr cm^{-3}}$. 
This then yields a significant 
decrease in the oscillation period at a few times $10^{15}$G. As we
will show in the next Section, this simple estimate is misleading, as
it completely ignores the magnetic field configuration. In practice,
the restoring force for the torsional oscillations will not be augmented
by the magnetic field tension equally throughout the star, but the influence
of the magnetic field on the oscillation will be highly dependent on the
correlation between the multipoles of the magnetic field configuration
and the quadrupole toroidal velocity field of the torsional mode.
For example, as we are showing in the next section, the radial component
of the magnetic field has a much weaker influence on the torsional modes
than other components. Thus, simple estimates as the one made above, are 
not sufficiently accurate for obtaining quantitative results that could 
be used for identifying normal modes in observations.

\section{A Special Case for the Magnetic Field}
\label{special}

In order to arrive at a one-dimensional boundary value problem, we consider
the following magnetic field configuration as a toy-model: 
\begin{eqnarray}
H^r&=&H^r(r)\neq0, \\ 
H^{\theta}&=&0,
\end{eqnarray}
For this case, Maxwell's equations reduce to 
\begin{equation}
H^r{}_{,r} + \left( \Lambda_{,r}+\frac{2}{r} \right) H^r=0.
\label{Maxspecial}
\end{equation}
The solutions admitted by (\ref{Maxspecial}) are of the form 
\begin{equation}
H^r=\frac{e^{-\Lambda}}{r^2} \mu_0,
\end{equation}
where $\mu_0$ is a constant. Although this form of the magnetic field
cannot be considered realistic, we will show that this toy-model can
be used for obtaining first estimates of the influence of the magnetic
field on the oscillation frequencies.  In our derivation, we assume
that the magnetic field has the above form only inside the solid crust
and not throughout the star.

Using (\ref{Maxspecial}), one can reduce the two-dimensional boundary
value problem (\ref{final}) to the following one-dimensional form:
\begin{eqnarray}
-\omega^2\left(\epsilon+p+H^2 \right)r^4e^{\Lambda-\Phi}Y &=&
 \left\{ \left[ \mu+ \left(e^\Lambda H^r \right)^2 \right]
e^{\Phi-\Lambda}r^4 Y_{,r} \right\}_{,r} 
+(\Lambda_{,r}-\Phi_{,r}) e^{\Phi+\Lambda} r^4 \left(H^r \right)^2 Y_{,r} 
\nn \\
&& - \left\{ (l+2)(l-1) \mu + r^2\left[ \left(\Phi_{,r} \right)^2 + \Phi_{,rr} 
\right) (H^r)^2 \right\} e^{\Phi+\Lambda}r^2 Y.
\label{wavespecial}
\end{eqnarray} 
This equation is a generalization of equation (66a) in Schumaker \&
Thorne (1983) to the case of the magnetic field configuration assumed
above, in the Cowling approximation. We note that both the influence
of the equilibrium magnetic field as well as the coupling between
torsional and magnetic field oscillations are included in our
derivation.

Comparing eqn. (\ref{wavespecial}) to the non-magnetized case, on sees
that the magnetic field enters through different terms. First, on the
l.h.s.  of (\ref{wavespecial}), the total energy density of the
hydromagnetic medium becomes $\epsilon + p + H^2$, as one would expect. On
the r.h.s., an important contribution of the magnetic field is
the augmentation of the shear modulus $\mu$ (in the term involving the
second spatial derivative of $Y$) by the magnetic field term $(e^\Lambda
H^r)^2$. This is the only magnetic field term on the r.h.s. that
survives in the Newtonian limit, which is in agreement with the
Newtonian results in Carroll et al. (1986) (see also Duncan, 1998).
The other two contributions of the magnetic field on the r.h.s. are
purely relativistic effects. Carroll et al. (1986) only looked at a
cylindrical model of the neutron star crust near the magnetic polar
cap, with a uniform magnetic field normal to the surface of the crust.
The one-dimensional boundary value problem considered in this section
is a generalization of Carroll et al.'s cylindrical Newtonian treatment 
to a spherical geometry and to general relativity.

Eigenvalues of the torsional modes in magnetized stars can be obtained
by solving (\ref{wavespecial}) as an eigenvalue problem with
zero-traction boundary conditions at the base of the crust and at the
surface of the star. For the oscillations of the magnetic field, we 
assume approximate boundary conditions, neglecting the oscillations
in the exterior of the solid crust. For the numerical solution of 
the above second-order eigenvalue problem, we introduced new variables 
$Y_1$ and $Y_2$ (see Carroll et al. (1986) and Leins (1994) for similar
definitions):
\begin{eqnarray}
Y_1&=&Y \\
Y_2&=&\left[\mu+\frac{ \left( e^\Lambda H^r \right)^2}{4 \pi} \right ] \frac{dY_1}{dr}
     e^{\Phi-\Lambda },
\end{eqnarray}
(note that we have restored the factor of $4\pi$ dividing the magnetic field).
This has the advantages of avoiding the numerical evaluation of the
derivative of the shear modulus $\mu$ (which is known in a 
tabular form only, for realistic equations of state) and that $Y_2$ vanishes
at the boundaries when zero-traction is assumed. With this definition,
the problem of finding the eigenfrequencies $\omega$ of torsional modes reduces
to solving the system of equations 
\begin{eqnarray}
\frac{dY_1}{dr}&=& \frac{Y_2}{ \mu+\frac{ \left( e^\Lambda H^r \right)^2}{4 \pi} 
} e^{\Lambda-\Phi}, \\
\frac{dY_2}{dr} &=& - \left[ \frac{4}{r} +  
    \frac{ (\Lambda_{,r}-\Phi_{,r}) (e^{\Lambda} H^r)^2}
     { 4 \pi\mu+ \left(e^\Lambda H^r \right)^2} \right]Y_2 \nn \\
  &&+\left\{ \left[(l+2)(l-1)\mu + r^2 \left(  \left(\frac{d\Phi}{dr} \right)^2
  +\frac{d^2\Phi}{dr^2} \right)\frac{ \left(H^r\right)^2}{4\pi} \right]
  \frac{e^{2\Phi}}{r^2} -\omega^2\left(\epsilon+P+\frac{H^2}{4\pi}\right)
   \right\} e^{\Lambda -\Phi} Y_1, \label{dY2}
\end{eqnarray}
with boundary conditions $Y_2=0$ at the surface and the base of the crust
and $Y_1$ being any finite number at the base of the crust. 

\begin{figure}
\centerline{\psfig{file=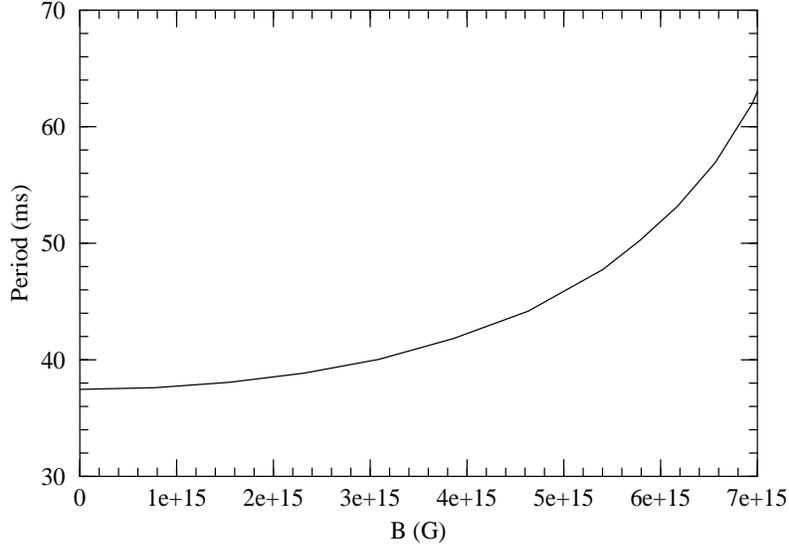,width=10.5cm,height=7.5cm}}
\caption{Oscillation period of the fundamental l=2 torsional mode as a 
function of the magnetic field strength, for a 1.4$M_\odot$ relativistic star and
for the simple magnetic field configuration considered in Section 
\ref{special}. The influence of the magnetic field is opposite to that 
predicted by simple estimates (see Section \ref{simple}), which 
demonstrates that the oscillation period is highly dependent on the given 
magnetic field configuration.}

\label{fig1}
\end{figure}

We construct an equilibrium TOV model, using the Wiringa et al. (1988)
equation of state (WFF3) matched to the Negele and Vautherin (1973)
equation of state at low densities. The star has a gravitational mass
of M=1.4 $M_\odot$ and a radius of $R=10.84$ km. We assume that the crust
extends from energy densities of $2.4 \times 10^{14} {\rm gr/cm^{-3}}$ down
to energy densities of $5\times 10^{10}{\rm gr/cm^{-3}}$, which corresponds
to radii from 9.86km to 10.65km (we truncate the crust at very low
densities to avoid numerical difficulties - this does not
significantly affect the obtained eigenfrequencies). We note that,
apart from the work of Leins (1994), the frequencies of even the
non-magnetized case have not been computed for realistic equations of
state in general relativity. In all previous Newtonian studies, the
eigenfrequency of the fundamental quadrupole torsional mode has been
found to be close to $~20$ms. In our example, this frequency is
$37$ms. This shows that the structure of a general relativistic star
and the relativistic perturbation equations yield torsional
frequencies very different from those predicted by the Newtonian
equations. For example, part of the difference between the Newtonian and
relativistic eigenfrequencies is due to the relativistic term
$e^{2\Phi}$, multiplying $\mu$ in equation (\ref{dY2}). For our model,
$e^{2\Phi}=0.59$ at the base of the crust. This effect increases the fundamental
torsional period by a factor of roughly $1/\sqrt{0.59}\simeq 1.3$. The 
remaining difference between our obtained $37$ms fundamental period and
the $17.32$ ms fundamental period for the most massive model considered in
McDermott et al. (1988) (model NS13T8) can be explained as follows: the
model considered in McDermott et al. (1988) has a mass of 1.326 $M_\odot$, 
similar to the mass of our present model, but the radius is only 7.853 km,
(compared to a radius of 10.65km for our model). The period of the
fundamental torsional mode has been shown to be roughly proportional to 
the star's radius (when the crust is thin) in Hansen and Cioffi (1980),
who estimate the $l=2$ period as $24.5 (R/10{\rm km})$ms. Using the
radius of our model and correcting for general relativity, the above
estimate becomes $34.5$ms, which is close to our numerically obtained
result of $37$ms. Thus, a general-relativistic version of
Hansen \& Cioffi's formula for the period of the fundamental torsional
mode is
\begin{equation}
P \simeq  34{\rm ms} \left( \frac{R}{10{\rm km}}\right),
\end{equation}
which should hold for $1.4 M_\odot$ models constructed with various realistic
equations of state, with the possible exception of strange star models 
with very thin crusts.

Figure 1 shows the period (in ms) of the fundamental quadrupole torsional
mode as a function of the magnetic field $B=H^r$. Due to the relativistic
factor which counteracts the shear modulus in (\ref{dY2}), the period is increased
by the magnetic field for values $B>2\times10^{15}$G. The change in the 
period does not follow the dependence predicted by the simple estimate 
(\ref{P2}) , which predicted that the magnetic field decreases the oscillation
period at a few times $10^{15}$G. 
We conclude that the eigenfrequency of
torsional modes in the presence of a strong magnetic field cannot be simply
estimated, assuming that the magnetic tension augments the shear modulus
in restoring the oscillation, isotropically. Instead, the degree to which
the magnetic field modifies the eigenfrequency is highly dependent on the 
magnetic field configuration and on the correlation of the latter with
the toroidal velocity field of the oscillation. 

In our example, the magnetic field has only a radial component. A careful
inspection of (\ref{dY2}) and (\ref{final}) reveals that the main term that
determines the eigenfrequency, $(l+2)(l-1)\mu$, is not augmented by the
radial component of the magnetic tension in the Newtonian limit, but only
by the azimuthal component. Thus, for a general configuration of the 
magnetic field, having both radial and azimuthal components, the numerical
results of Figure 1 can be considered as one limiting case of the influence
of the magnetic field on the torsional modes, while the isotropic
simple estimate (\ref{P2}) would be another limiting case. We
expect that the frequencies of torsional modes for realistic magnetic
field configurations will have values within these limits.

\section{Discussion}

We initiate a study of torsional oscillations in relativistic stars
possessing a strong magnetic field, based on the treatment of 
wave-propagation in hydromagnetic media in general relativity by
Papadopoulos \& Esposito (1982). We derive the linear perturbation
equations governing torsional oscillations in a magnetized star in the
relativistic Cowling approximation, neglecting the deformation of the
equilibrium structure due to the presence of a magnetic field and
variations of the metric. The perturbation equations are
two-dimensional in the case of a general axisymmetric magnetic field
configuration.  Simplified, one-dimensional equations are derived for
a limiting case of the magnetic field configuration. This allows first
estimates for the change in the mode-frequencies, due to the magnetic
field, to be obtained. The origin of this change is both due to the
equilibrium energy density of the magnetic field and due to the
back-reaction of the magnetic field oscillations on the oscillations
of the solid crust. Our results are substantially different from a
simple isotropic estimate, showing that the influence of the 
magnetic field on the frequency of torsional oscillations is highly
dependent both on relativistic effects and the structure of the magnetic field.

The study of torsional oscillations is motivated by the prospect that
they could become observable in gravitational waves or in high-energy
radiation after a crust fracture is initiated by a strong magnetic
field. Work by Duncan (1998) suggests that in this scenario the
torsional modes of the crust will be the dominant mode of oscillation,
as they do not couple strongly to density variations. In the latter
reference it is suggested that an observed periodicity of 21ms in the
March 5 event could in fact be the signature of the fundamental
quadrupole torsional oscillation of the neutron star's crust. For this
to happen, the star must have a very strong magnetic field, since, as
we show, this mode has a longer period of 37ms in a typical relativistic
nonmagnetized star (contrary to the 20ms estimate derived in previous
Newtonian studies).

The correct identification of such periodicities with a specific
normal mode requires the computation of mode-frequencies in the
presence of a strong magnetic field in full general relativity. A
successful identification will significantly constrain the properties
of the high-energy equation of state in relativistic stars.  As no
other mode of oscillation has been identified in neutron stars, to
date, torsional modes could be the oscillations to initiate the field
of observational neutron-star asteroseismology. Especially a
combination of mode-identifications obtained in high-energy
radiation observations with  mode-identifications in
gravitational radiation detections (see Andersson and Kokkotas 1996,
Kokkotas, Apostolatos and Andersson 2001) should yield invaluable
information on the properties and internal composition of relativistic
stars.

In future work, we plan to present detailed numerical results for
various realistic neutron star equations of state and for various
axisymmetric magnetic field configurations, by numerically solving the
full two-dimensional eigenvalue problem.

\section{Acknowledgments}

We are grateful to Nils Andersson, Valeria Ferrari, Wlodek Kluzniak, Kostas D.
Kokkotas, Luciano Rezzolla, Johannes Ruoff and S. Yoshida for useful
discussions and comments on the manuscript. This research was
supported in part by the EU Programme 'Improving the Human Research Potential
and the Socio-Economic Knowledge Base' (Research and Training Network Contract
HPRN-CT-2000-00137).


\end{document}